\documentclass[11pt]{article}
\usepackage{moriond,epsfig}

\def\Journal#1#2#3#4{{#1} {\bf #2}, #3 (#4)}

\def\NIMA{{\em Nucl. Instrum. Methods} A}

\def\PLB{{\em Phys. Lett.}  B}

\def\PRD{{\em Phys. Rev.} D}

\def\APP{\em Astropart. Phys.}
\def\ApJ{\em Astrophys. J.}

\def\DSR1{{\em Deep-Sea Res.} I}

\def\proc{{\em these proceedings}}

\def\be{\begin{equation}}
\def\ee{\end{equation}}
\def\bea{\begin{eqnarray}}
\def\eea{\end{eqnarray}}

\newcommand{\gradi}{\ensuremath{^\circ}}

\begin{document}
\vspace*{4cm}
%\title{ENERGY SPECTRUM OF ATMOSPHERIC  $\nu_\mu$ AND LIMITS  ON  \\ DIFFUSE FLUXES OF COSMIC NEUTRINOS WITH ANTARES}
\title{UPDATED LIMITS  ON   DIFFUSE FLUXES OF \\ COSMIC NEUTRINOS WITH 2008-2011 ANTARES DATA}

\author{SIMONE BIAGI   on behalf of the ANTARES Collaboration}

\address{Dipartimento di Fisica dell'Universit\`a and INFN Sezione di Bologna, \\
Viale Berti Pichat 6/2, 40127 Bologna, Italy}

\maketitle\abstracts{
The ANTARES neutrino telescope detects the Cherenkov radiation emitted along the path of charged particles produced in neutrino interactions. % inside or in the vicinity of the detector. 
ANTARES is sensitive to all flavors even though it is optimized for muon neutrinos. Several algorithms estimating the deposited energy in the active volume of the detector have been developed and applied to the reconstruction of the primary neutrino energy -- this allows to improve the search for a diffuse flux of astrophysical neutrinos. 
The search for a diffuse flux of cosmic neutrinos at very high energies (E$_\nu > 30$  TeV) is updated  using 4 years of data  with the full detector.
}

%%%%%%%%%%%%%%%%%%%%%%%%%%%%%%%%%%%%%%%%
%\section{Introduction} \label{sec:intro}
\section{ANTARES and Diffuse Cosmic Neutrinos} \label{sec:main}

\begin{figure}[t!]
\centering
\psfig{figure=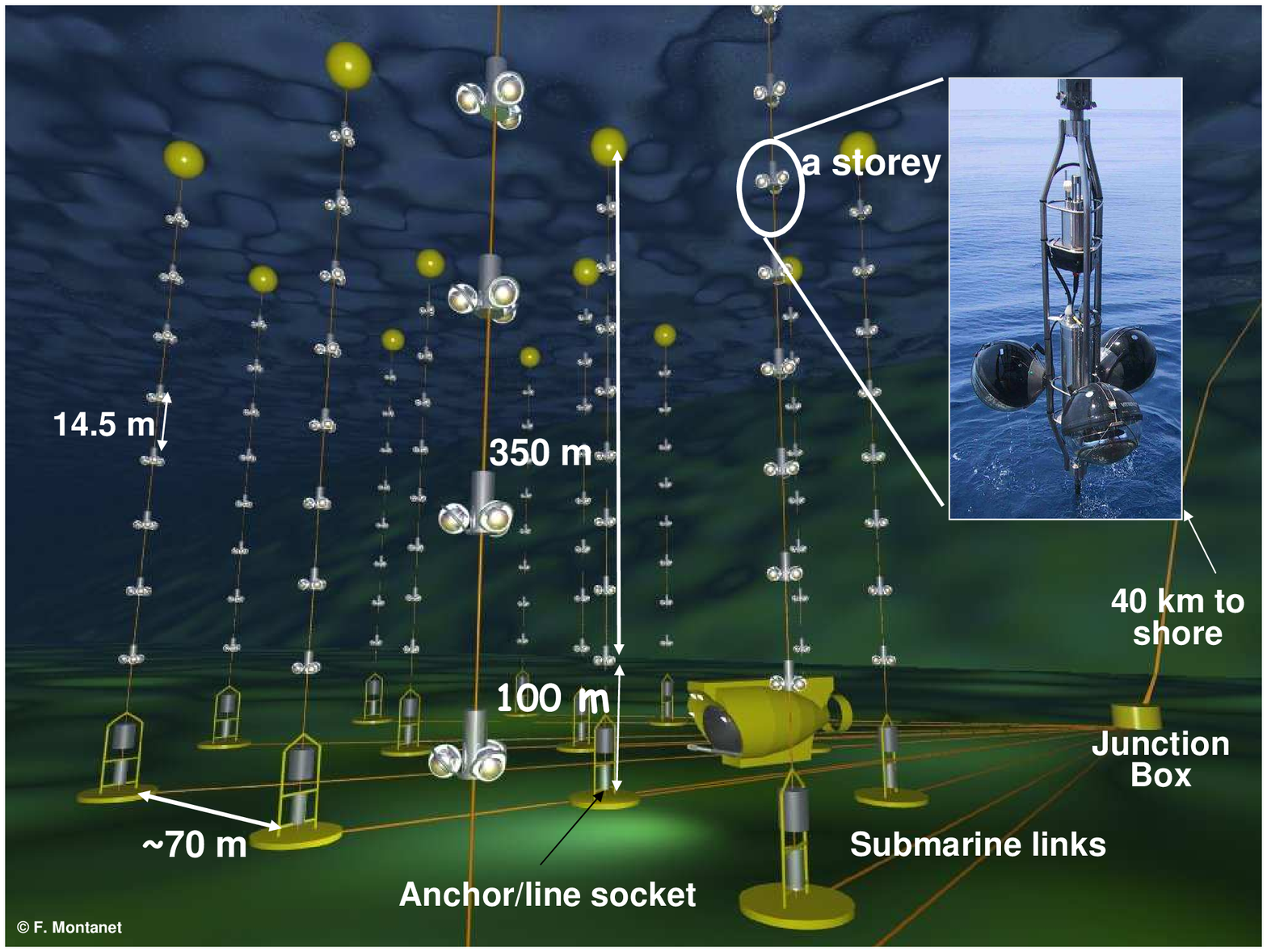,width=0.45\textwidth} 
\psfig{figure=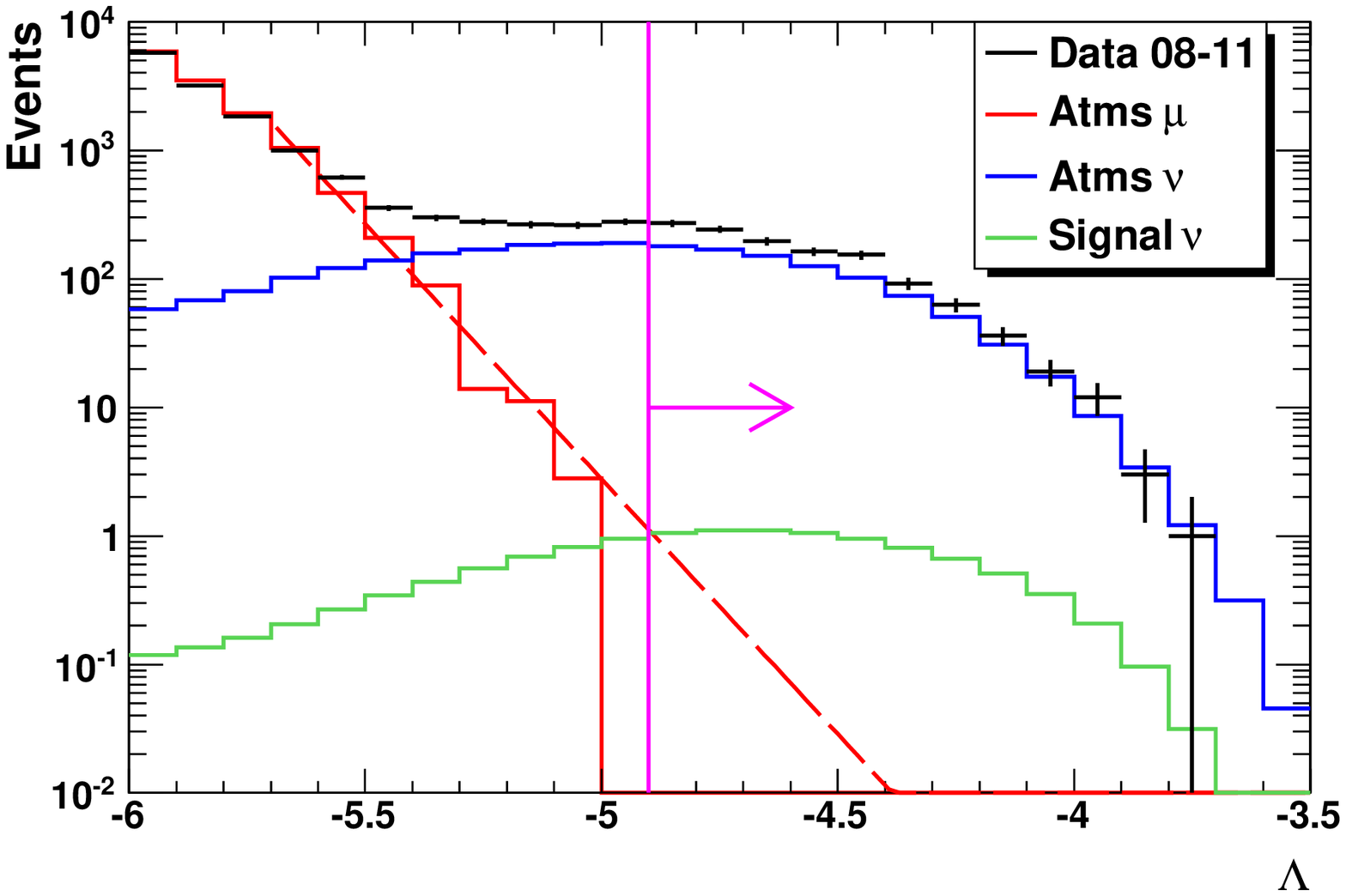,width=0.54\textwidth} 
\caption{Left: A schematic view of ANTARES. Right:  Data and MC events with an equivalent live-time of 885 days as a function of $\Lambda$, selected with $\theta > 90\gradi$, $\beta < 0.5$, $N_{hit} > 35$.  $\Lambda$ is negative and takes values closer  to zero for well reconstructed tracks. 
%The cut at $\Lambda > -4.9$ is show as a pink line. The tail of the MUPAGE distribution is fitted with an exponential. After the cut, the estimated muon contamination is the 0.15\% (1.3 $\mu$).  
Red represents  atmospheric muons, blue  the  conventional atmospheric neutrinos, green   the cosmic signal $\Phi_{test}\propto E^{-2}$, points are data. 
%Due to the simulated live-time of one third respect to data, 
The cut at $\Lambda > 4.9$ (pink vertical line) is adjusted according to an exponential fit  of the muon distribution
 that reduces the  muon background to   0.15\% (1.3 $\mu$) of the total.  }
\label{fig:lambda}
\end{figure}

The ANTARES neutrino telescope~\cite{antares} is taking data in its final configuration since 2008, see Fig. \ref{fig:lambda} (left). The main physics subject is the search for cosmic sources of neutrinos, even though  several results have been obtained in other topics, e.g. neutrino oscillations, searches for dark matter  and exotics (monopoles, nuclearites), oceanography and marine biology~\cite{mangano}. 
%Recent results have been presented  by  Mangano~\cite{mangano} during this conference.

An update %of the analysis devoted to the search for a diffuse flux of cosmic neutrinos already published in J.A. Aguilar {\it et al.}~\cite{diffusi09} is presented.
of the analysis on the search for a diffuse flux of cosmic neutrinos~\cite{diffusi09} %where  two years of data were analyzed is presented.
with two years of data   is presented.
%In the cited paper, two years of data were analyzed. % and a world-record upper limit on diffuse fluxes obtained. 
Two years more of data are added  to the measurement.
In the meanwhile, our knowledge of the detector has improved and better  Monte Carlo (MC) simulations have been made available, allowing to use a larger data fraction for analyses, with less requirements on the data quality.
The equivalent live-time is 885 days now, about a factor three larger than for previous analysis.
%Two years more of data are added here to the measurement  and the equivalent live-time is 885 days now, about a factor three larger than previous analysis.
%Now, more data are available. 

%Muons are produced  in the proximity  of the instrumented volume  by muon neutrinos and antineutrinos.  
%In the following we refer to $\nu_\mu$ and $\overline{\nu}_\mu$ as ``muon neutrinos'', because  the sign of the charged muon  is indistinguishable.

%%%%%%%%%%%%%%%%%%%%%%%%%%%%%%%%%%%%%%%%
%\section{Background treatment} \label{sec:background}

An isotropic diffuse flux of neutrinos generated by  extragalactic sources in the Universe is expected. 
Atmospheric neutrinos with an energy spectrum  $\propto E^{-3.7}_\nu$ represent an irreducible background   
in  the search for a diffuse flux of cosmic $\nu$ and the signal 
%and the contribution from a diffuse flux of cosmic neutrinos 
might be seen as an excess in the high energy region of the spectrum. 
A harder  spectrum, proportional to E$^{-2}$, is expected for $\nu$ of astrophysical origin; above an unknown value of the critical energy E$_\nu^c$ (which depends on the absolute normalization of the cosmic $\nu$ flux) astrophysical neutrinos should exceed those  of atmospheric origin. The discrimination can be done on the basis of the visible energy of  muons generated by neutrinos.
In the following we refer to $\nu_\mu$ and $\overline{\nu}_\mu$ as ``muon neutrinos'', because  the sign of the charged muon  is indistinguishable.

IceCube results on the search for diffuse neutrinos %were presented  at the Neutrino 2012 conference  using IC59 data~\cite{sullivan}. 
using IC59 data~\cite{sullivan} show 
%No significant excess of cosmic neutrinos with respect to the background expectations was found, hence an upper limit at 90\% c.l. was settled:
no significant excess  with respect to the background expectations, hence an upper limit at 90\% confidence level  was derived:
\begin{equation}\label{eq:ic59_lim}
E^2 \Phi_{\nu_\mu} = 1.44 \cdot 10^{-8}   \  \mathrm{GeV\ cm^{-2}\ s^{-1}\ sr^{-1}}   .
\end{equation}
%
%The detector sensitivity to diffuse fluxes was quoted equal to $7.5 \cdot 10^{-9}$ GeV cm$^{-2}$ s$^{-1}$ sr$^{-1}$, a factor two lower than the limit. Even if this result is statistically compatible with zero at 1.8 $\sigma$, there is a soft indication of the presence of high energy astrophysical neutrinos.
The IC59 sensitivity to diffuse fluxes is a factor two lower than the limit. Even if this result is statistically compatible with zero at 1.8 $\sigma$, there is a soft indication of the presence of high energy astrophysical neutrinos.

A test $\nu_\mu$ signal with a flux proportional to E$^{-2}$ and normalized at
\begin{equation}\label{eq:signal}
E^2 \Phi_{test} =  10^{-8}   \  \mathrm{GeV\ cm^{-2}\ s^{-1}\ sr^{-1}}   
\end{equation}
is used in this analysis. The normalization is arbitrary and does not affect the result of the following cut optimization.
The entire procedure and the optimization of all the cuts was done using only the 10\% of available data (blind analysis) and the full MC set, in order to avoid any bias.

%The first step  is to remove as much as possible the contamination due to atmospheric muons. 
%Furthermore, d
Down-going atmospheric muons reconstructed as up-going can mimic high-energy neutrino induced muons. 
In fact, atmospheric muons reach the detector in ``bundles'' of particles with a large multiplicity. % (up to $\sim$1000), releasing a lot of energy in the sea water. 
The main effect is the production of  a large amount of hits on the PhotoMultiplier Tubes (PMTs) -- the signature of high energy events. 
A cut on the quality of  reconstructed tracks is defined to keep under control the atmospheric muon background.
%This analysis is to find a small excess of neutrino events in the high energy tale of the event distribution. Such a kind of background is greatly dangerous and can introduce errors in final result.

The algorithm that reconstructs the muon direction~\cite{lambda} uses as input  the time and position information of hits produced by the photons recorded by the PMTs, and gives as output the direction of the muon (zenith and azimuth), a quality parameter ($\Lambda$), the estimation of the angular resolution ($\beta$), and the number of hits correlated with the track ($N_{hit}$).
An a priori cut on the reconstructed zenith angle is applied
($\theta > 90 \gradi$).
%%
%\begin{equation}\label{eq:theta}
%\theta > 90 \gradi
%\end{equation}
%%
%to ensure the removal of down-going muons. 
A combined cut on the  three parameters given by the reconstruction algorithm ($\Lambda$, $\beta$, $N_{hit}$) has been optimized to  reduce the muon background  %at the level of  one event 
maximizing the total  number of signal events. 
%Using the entire MC sample, the weighted number of MUPAGE and E$^{-2}$  $\nu$ events is evaluated changing the cut on the three parameters and summary tables %

First level cuts are defined as the combination of:
\begin{equation}\label{eq:FLC}
%\begin{array}{lcr}  
%\theta  &>& 90\gradi , &  & 
%\Lambda &>& -4.9 \\
%\beta &<& 0.5 & & 
%N_{hit} &>& 35 \\
\Lambda > -4.9 , \ \ 
\beta < 0.5 , \ \ 
N_{hit} > 35 .
%\end{array}	
\end{equation}
%
%The set of cuts that accomplish this condition is $\Lambda>-4$, $\beta<0.5$, $N_{hit}>35$. 
Due to the reduced statistics in the muon MC sample (one third of the equivalent live-time in data) the cut on $\Lambda$  is adjusted   according to the fit shown in Fig. \ref{fig:lambda} (right). The first level cuts allow to reduce the muon background at the level of 1.3 event   (885  days).

%%%%%%%%%%%%%%%%%%%%%%%%%%%%%%%%%%%%%%%%
%\section{Energy estimation} \label{sec:energy}

%Assuming negligible the contribution of atmospheric muons after the FLCs, 
After the cuts of eq. \ref{eq:FLC} 
the %main source of  
prevailing % dominant 
background for   cosmic neutrinos is   due to atmospheric neutrinos, which are expected to dominate below the critical energy E$_\nu^c$.  
The neutrino energy cannot be directly measured %. And the neutrino induced muon energy cannot be directly  measured too
and the neutrino induced muons are observed in a limited interval of their range,
due to the limited size of  ANTARES. 
%In fact, muons with an energy of 1 TeV can travel several kilometers in water: only part of the deposited energy in sea water is visible from the ANTARES Optical Modules (OM). %In other words, at the level of the  energies considered here the fraction of contained events is negligible.
An estimate  of the muon energy can be done through the measurement of some observables related with the muon energy loss in water. In fact, at energies higher than $\sim500$ GeV, the energy loss is   proportional to the energy of the muon.
%The algorithm described in~\cite{dEdX} uses as inputs some parameters related with energy, like the number of hits and the total charge on the OMs, together with some geometrical factors, like the detector efficiency and the estimated track length. A detailed study of the estimator properties can be found in the cited internal note, while a data-MC comparison of the intermediate parameters of the estimator  after unblinding is shown in Fig. \ref{fig:parametri}.
An energy estimator~\cite{fabian}, $\rho$, is defined through an approximation of  the total muon energy deposited in the detector along its path,  
%$\Delta E / \Delta x$,  and it can be written as:
%%
%\begin{equation} \label{eq:dEdX}
%\Delta E / \Delta x \propto \rho = \frac{\sum\limits^{N_{hit}}_{i=1} Q_i}{\epsilon(\overrightarrow{x})} \cdot \frac{1}{L_\mu (\overrightarrow{x})}
%\end{equation}
%%
%where $\rho$ is the estimator itself, $Q_i$ is the charge in p.e. recorded by PMTs, $\overrightarrow{x}$ is the reconstructed muon direction, $\epsilon$ is the detector efficiency for that particular $\overrightarrow{x}$, $L_\mu$ is the geometrical track length within the sensitive volume.
 $\Delta E / \Delta x \propto \rho(Q_i, \overrightarrow{x}, L_\mu, \epsilon)$,  and it is a function of   the total number of photoelectrons   recorded by PMTs ($Q_i$),  the reconstructed muon direction ($\overrightarrow{x}$), the geometrical track length within the sensitive volume ($L_\mu$), and  the detector efficiency ($\epsilon$).

\begin{figure}[t!]
\centering
\psfig{figure=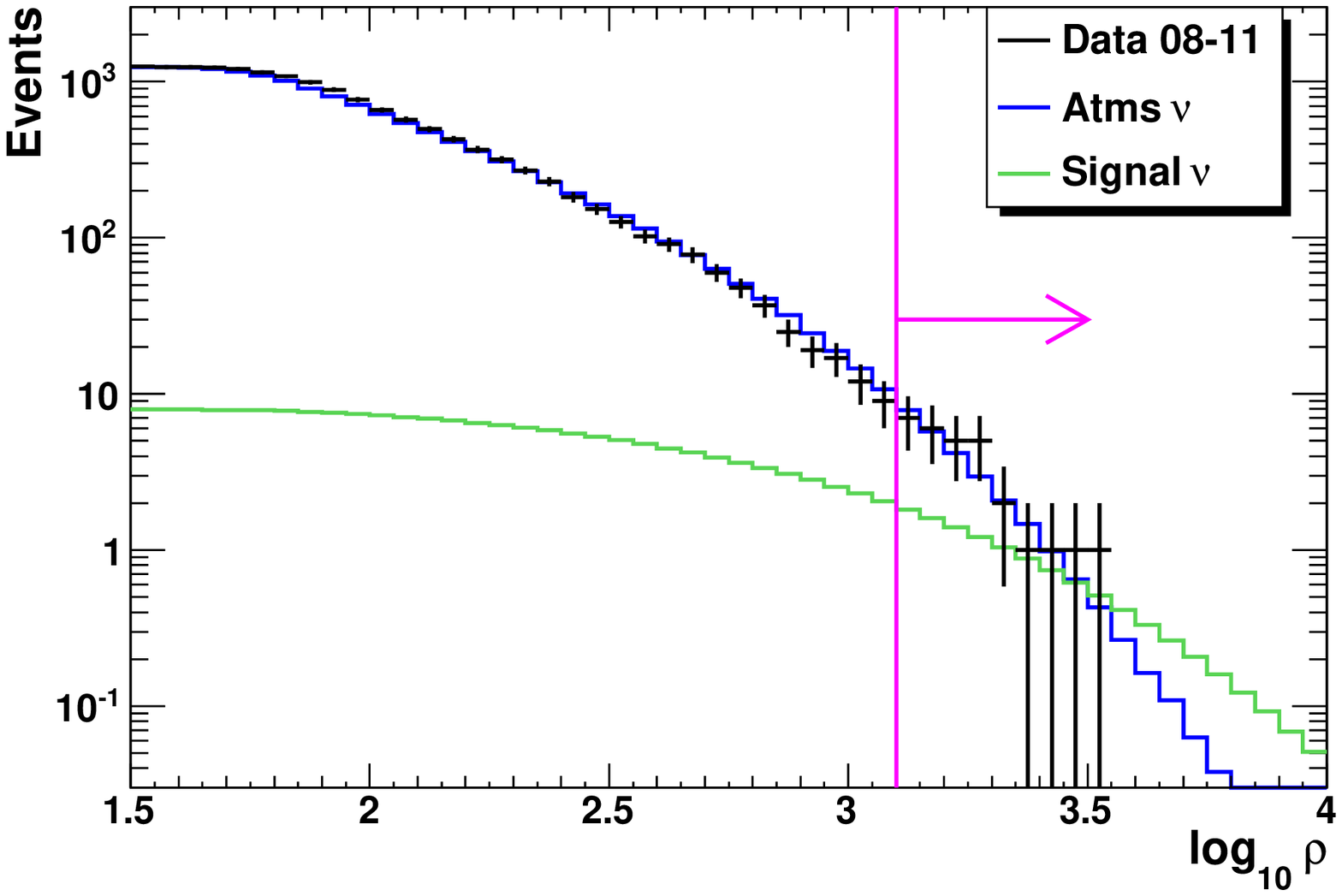,width=0.496\textwidth}
\psfig{figure=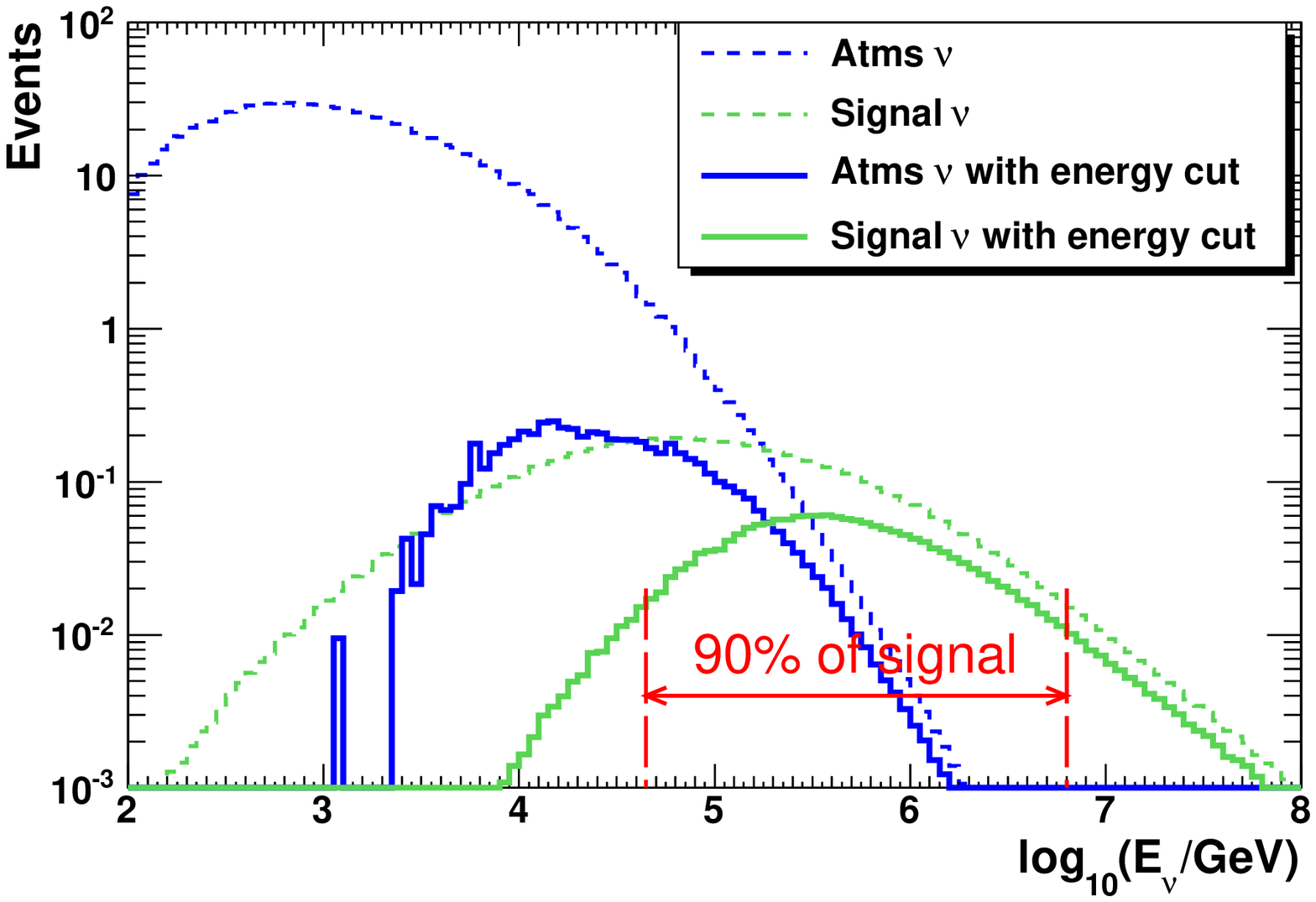,width=0.496\textwidth}
\caption{Left: Inverse cumulative distributions of data (crosses) and Monte Carlo as a function of the energy estimator $\rho$.  The blue histogram represents the atmospheric Bartol $\nu_\mu$    normalized to the total number of data events, without any prompt contributions; the green histogram the cosmic neutrinos with the test flux of eq. \ref{eq:signal}; the pink vertical line shows the position of the cut which minimizes the MRF.
Right: Distribution of atmospheric and signal $\nu_\mu$ as a function of the true energy obtained from MC, before and after the energy cut $\log \rho > 3.1$.  The region that contains the 90\% of signal is highlighted. }
\label{fig:energy}
\end{figure}

%The discrimination between the cosmic neutrino signal and the atmospheric neutrino energy is done on the basis of the energy. 
The atmospheric neutrinos are simulated  according  to the conventional    ``Bartol''  flux~\cite{bartol}, 
while the signal is taken assuming the flux of eq. \ref{eq:signal}.  
The inverse cumulative  distributions  of expected neutrinos as a function of $\log \rho$ is shown  in Fig. \ref{fig:energy} (left).
%The energy estimator $\rho$ is used to define the cut which gives the best sensitivity applying the Model Rejection Factor (MRF) procedure~\cite{mrf}. 
The energy estimator $\rho$ is used in the Model Rejection Factor (MRF) procedure~\cite{mrf} to define the cut which gives the best sensitivity. 
%With the Feldman\&Cousins statistics~\cite{feldman} it is possible to derive the MRF as a function of the cut on $\rho$, that represents the sensitivity of the ANTARES detector at a 90\% confidence level  in units of 10$^{-8}$ GeV cm$^{-2}$ s$^{-1}$ sr$^{-1}$ as a function of the chosen cut.
The MRF as a function of $\rho$ was computed through pseudo-experiments using the Feldman\&Cousins statistics~\cite{feldman} at 90\% confidence level. 
The minimum %(MRF=3.0) 
occurs selecting events with  $\log \rho > 3.1$ and corresponds to a sensitivity:
\begin{equation}\label{eq:sensitivity}
\Phi_{sens} =  3.0 \cdot 10^{-8}  E^{-2}    \  \mathrm{GeV\ cm^{-2}\ s^{-1}\ sr^{-1}}    .
\end{equation}

%%%%%%%%%%%%%%%%%%%%%%%%%%%%%%%%%%%%%%%%
%\section{Upper limit} \label{sec:limit}
\section{Results} \label{sec:results}

%\begin{table}[b]
%\label{tab:values}
%\caption{Number of events for data and MC after the First Level Cuts  for different energy cuts.  The Bartol flux has a deficit of 27\% respect to data and it is normalized to the total number of neutrinos in data. The signal  is from the test normalization  of eq. \ref{eq:signal}. The last row is the cut obtained with the MRF procedure. }
%\vspace{0.4cm}
%\begin{center}
%\begin{tabular}{c|cccc}
%\textbf{   } & \textbf{ Data } & \textbf{ Bartol }  & \textbf{ MC (norm.) }  & \textbf{ E$^{-2}$ signal }  \\
%\hline 
%total                  & 1253 & 914.8 & 1253   & 7.9  \\
%$\log \rho > 2.0$ & 657  & 453.0 & 620.5  & 7.3  \\
%$\log \rho > 2.5$ & 126  & 100.6 & 137.8  & 5.0  \\
%$\log \rho > 3.0$ & 12   & 10.6 & 14.5      & 2.3  \\
%\hline 
%$\log \rho > 3.1$ &  7    & 5.7 & 7.9      & 1.8  \\
%\hline 
%\end{tabular}
%\end{center}
%\end{table}

Applying the first level cuts  to the data sample, the conventional atmospheric neutrinos from MC simulations  show a 27\% deficit  with respect to the observed data events. 
This is well within the systematic uncertainties on the theoretical expectation at these energies; the  $\nu$  background is then normalized to  the  data.
After normalization,  7.9 atmospheric events are expected for $\log\rho > 3.1$, and 1.8 signal events from the test flux (eq. \ref{eq:signal}).

%After unblinding the final data sample, 7 neutrino events are found in the high energy region. 
After unblinding the high energy region,  7 neutrino events are found in the full data sample.
Fig.  \ref{fig:energy} (left) shows the inverse cumulative distribution as a function of the energy estimator $\rho$ for  %the expected atmospheric neutrinos, the signal from the test flux (eq. \ref{eq:signal}) and data.
data and MC.
The pink line at $\log\rho = 3.1$ shows the cut value which minimizes the MRF.

The effects of   systematics  is considered  in the calculation of the upper limit.
The evaluation of the  systematic errors on the background  is taken from the normalization factor applied to Monte Carlo, giving a $\pm27\%$ effect on the predicted 7.9 events. This factor includes the systematics about the knowledge of the detector plus theoretical uncertainties on the conventional neutrino flux. 
Concerning the signal, an assumption is done on the flux shape and the absolute  normalization does not influence the resulting upper limit. 
The variation   in the number of expected signal events depends on the detector efficiencies only. %; some critical parameters were changed in  MC simulations to evaluate  the signal expectations. 
Some critical parameters were changed in  the  MC simulations to evaluate  the signal expectations: absorption length of light in water ($\pm 10\%$), PMT quantum  efficiency ($\pm 10\%$),  PMT angular acceptance.
%The following parameters was tested:
%%
%\begin{itemize}
%\item absorption length $\pm 10\%$;
%\item PMT efficiency $\pm 10\%$;
%\item 2 different models of PMT angular acceptance (respect to standard MC).
%\end{itemize}
%%
The effect of the systematic uncertainties  is to change the expected 1.8 signal events by $\pm 14\%$.

%The corresponding upper limit is obtained including the effects of the systematic uncertainties discussed in \S \ref{uncertainties}.
%Their variations change the number of expected signal events by $\pm 14\%$. The normalization factor from data and Monte Carlo are assumed as systematic uncertainty for the background, giving a $\pm27\%$ effect on the predicted 7.9 events.

Using the method described in  Conrad {\it et al.}~\cite{conrad}, the upper limit at 90\% confidence level is:
\begin{equation} \label{eq:limit}
E^2 \Phi_{90\%} = 3.2 \cdot 10^{-8}   \  \mathrm{GeV\ cm^{-2}\ s^{-1}\ sr^{-1}}   
\end{equation}
The central 90\% of the signal is found in the neutrino energy range from 45 TeV to 6.3 PeV --~see Fig. \ref{fig:energy}  (right). 
The interval is  the region containing 90\% of the signal    from MC simulations.

\begin{figure}[t!]
\centering
\psfig{figure=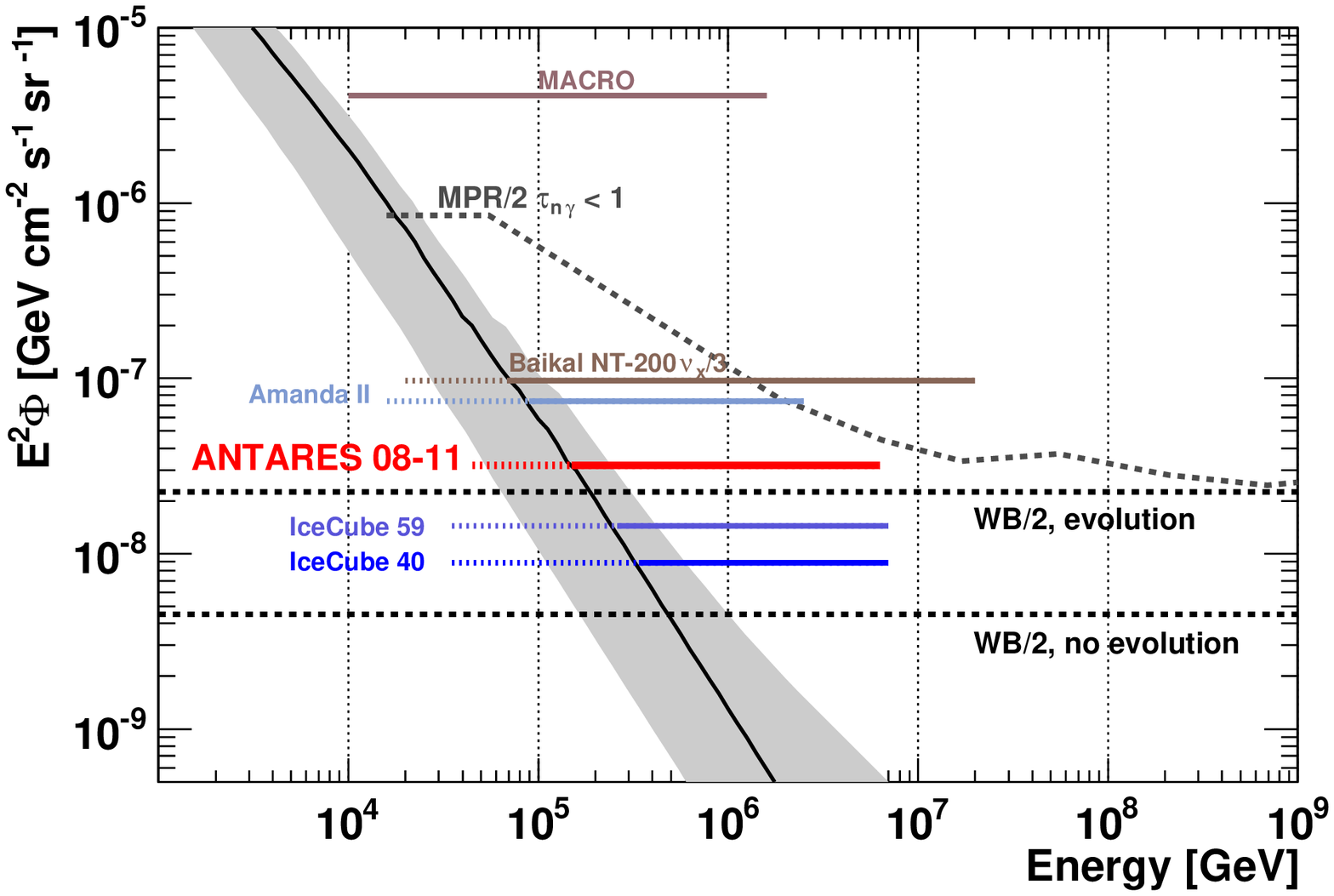,width=0.496\textwidth} 
\psfig{figure=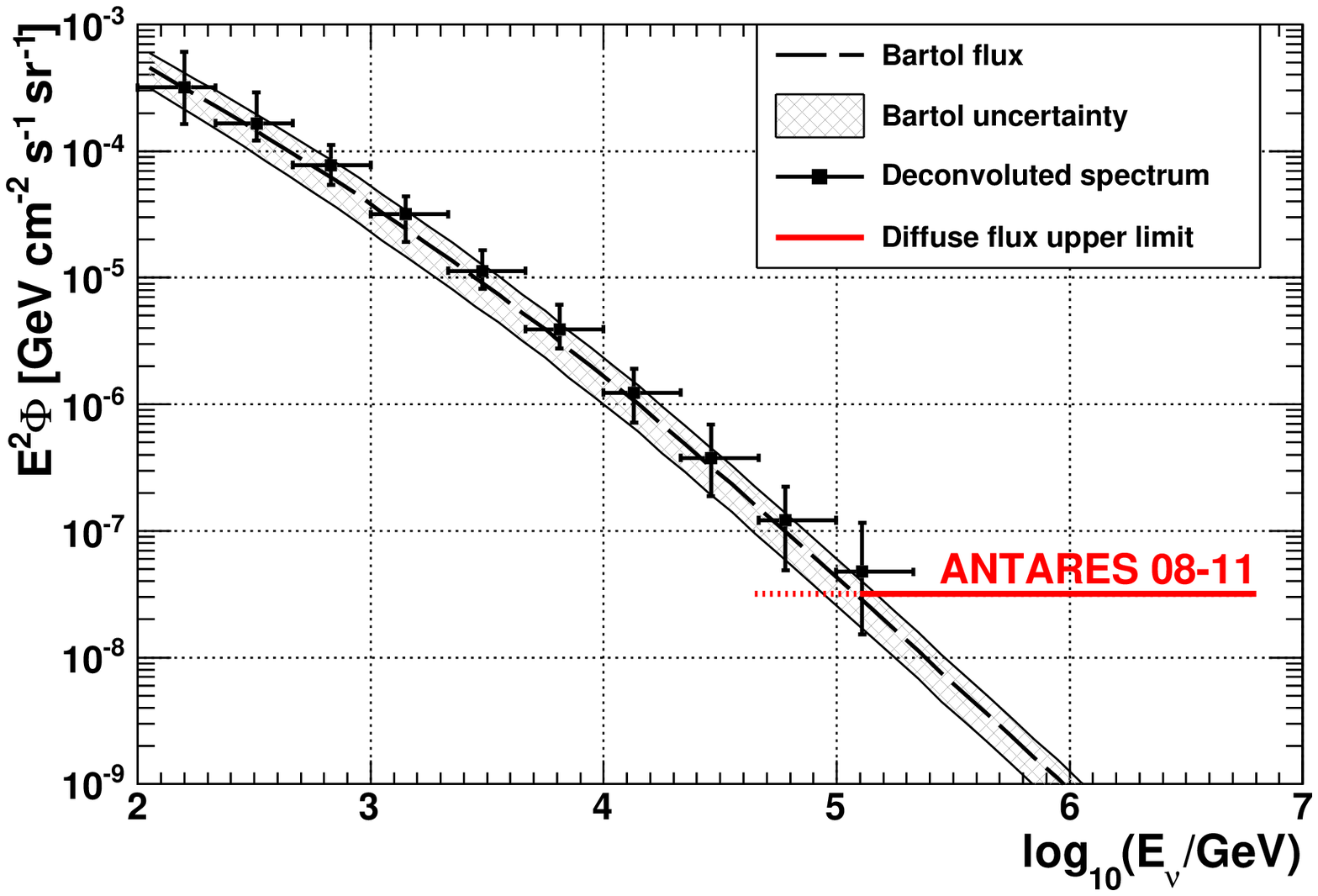,width=0.496\textwidth}
\caption{Left: The ANTARES 08-11 upper limit is compared with limits from other experiments and theoretical models. The gray band represents  the conventional Bartol flux from vertical to horizontal events.
Right: The  upper limit on diffuse fluxes  and the unfolded atmospheric neutrino energy spectrum  are shown for comparison in the same picture, together with the expectations  from the Bartol flux.}
\label{fig:limit}
\end{figure}

%%%%%%%%%%%%%%%%%%%%%%%%%%%%%%%%%%%%%%%%
\section{Conclusions} \label{sec:conclusions}

ANTARES data taken in the years 2008-2011 were analyzed to search for a diffuse cosmic neutrino signal. The whole period corresponds to 855 days of equivalent  live-time. Using an estimator  of the muon energy loss in sea water, no excess of events is found  with respect to the atmospheric neutrino flux hence an upper limit at     90\% c.l.  is obtained.  This result is compared with other experiments and theoretical expectations in Fig. \ref{fig:limit} (left) -- see~\cite{diffusi09} for references. The same data set was analyzed to unfold  the atmospheric neutrino energy spectrum~\cite{fusco}; Fig. \ref{fig:limit} (right) shows the combination of both results together with the conventional Bartol flux.


\section*{References}
\begin{thebibliography}{99}

\bibitem{antares} M. Ageron {\it et al.}, \Journal{\NIMA}{656}{11-38}{2011}.
\bibitem{mangano} S. Mangano, \proc.

\bibitem{diffusi09} J.A. Aguilar {\it et al.}, \Journal{\PLB}{696}{16-22}{2011}.
\bibitem{sullivan} G. Sullivan, {\em arXiv:}  {\bf 1210.4195} [astro-ph.HE].

\bibitem{lambda} S. Adri\'an-Mart\'inez  {\it et al.},  \Journal{\ApJ}{760}{53}{2012}.
\bibitem{fabian} F. Sch\"ussler,  {\it Proceedings for the 2013 ICRC}, ID 421 (2013).

\bibitem{bartol} V. Agrawal {\it et al.}, \Journal{\PRD}{53}{1314-1323}{1996}.
\bibitem{mrf} G.C. Hill and K. Rawlins, \Journal{\APP}{19}{393-402}{2003}.
\bibitem{feldman} G.J. Feldman and R.D. Cousins, \Journal{\PRD}{57}{3873-3889}{1998}.
\bibitem{conrad} J. Conrad {\it et al.}, \Journal{\PRD}{67}{012002}{2003}.

\bibitem{fusco} L.A. Fusco, {\it Proceedings for the 2013 ICRC}, ID 636 (2013).
\end{thebibliography}
\end{document}